\documentclass{PoS}
\usepackage{amsfonts,bm,mathbbol}
\usepackage{graphicx}
\usepackage{pstricks}

\newcommand{\be}{\begin{equation}}
\newcommand{\bea}{\begin{eqnarray}}
\newcommand{\ee}{\end{equation}}
\newcommand{\eea}{\end{eqnarray}}

\def\1eq#1{Eq.~(\ref{#1})}
\def\2eqs#1#2{Eqs.~(\ref{#1}) and~(\ref{#2})}
\def\3eqs#1#2#3{Eqs.~(\ref{#1}), (\ref{#2}) and~(\ref{#3})}
\def\4eqs#1#2#3#4{Eqs.~(\ref{#1}), (\ref{#2}), (\ref{#3}) and~(\ref{#4})}

\def\ie{{\it i.e.}, }

\title{The impact of the ghost-gluon vertex on the ghost  Schwinger-Dyson equations}

\ShortTitle{The impact of the ghost-gluon vertex on the ghost  Schwinger-Dyson equations}

\author{\speaker{A.~C.~Aguilar}\\
       University of Campinas - UNICAMP,\\ Institute of Physics ``Gleb Wataghin'',
13083-859 Campinas, SP, Brazil\\
        E-mail: \email{aguilar@ifi.unicamp.br}}

\abstract{We derive an approximate dynamical equation for the form-factor of the
ghost-gluon vertex that contributes to the Schwinger-Dyson equation of
the ghost  dressing function in  the Landau gauge.  In  particular, we
consider the  ``one-loop dressed'' approximation  of the corresponding
equation  governing the  evolution  of the  ghost-gluon vertex,  using
fully  dressed propagators  and  tree-level vertices  in the  relevant
diagrams.    Within   this   approximation,   we   then   compute   the
aforementioned form  factor for two  special kinematic configurations,
namely the \emph{soft  gluon limit}, in which the  momentum carried by
the gluon  leg is zero  , and the  \emph{soft ghost limit},  where the
momentum of the anti-ghost leg vanishes.  The results obtained display
a considerable departure from the  tree-level value, and are in rather
good agreement with available lattice data.  We next solve numerically
the  coupled system  formed  by  the equation  of  the ghost  dressing
function and  that of the  the vertex form  factor, in the  soft ghost
limit.  Our  results  demonstrate  clearly  that  the  nonperturbative
contribution  from the  ghost-gluon  vertex accounts  for the  missing
strength  in the  kernel  of the  ghost  equation, and  allows for  an
impressive coincidence  with the lattice results, without  the need to
artificially enhance the coupling constant of the theory.}

\FullConference{QCD-TNT-III-From quarks and gluons to hadronic matter: A bridge too far?,\\
		2-6 September, 2013\\
		European Centre for Theoretical Studies in Nuclear Physics and Related Areas (ECT*), Villazzano, Trento (Italy)}

\begin{document}

\section{Introduction}

A qualitative and quantitative understanding of fundamental Green's function  in 
the infrared (IR) sector constitutes a long-standing challenge in QCD.  In the last few  years, our knowledge of the QCD low energy regime has advanced considerably, due to a systematic efforts obtained through various non-perturbative methods,  
such as  by  lattice simulations~\cite{Cucchieri:2007md,
Cucchieri:2003di,Bogolubsky:2007ud,Oliveira:2008uf,arXiv:0912.0437,Bowman:2007du},
Schwinger-Dyson  equations (SDEs)~\cite{Aguilar:2008xm, Binosi:2009qm, Boucaud:2008ji,Aguilar:2009nf,Aguilar:2010cn}, 
functional methods~\cite{Braun:2007bx,Szczepaniak:2010fe,Quandt:2013wna},
and algebraic  techniques~\cite{Zwanziger:1993dh,Dudal:2008sp,Kondo:2011ab}. On the level of the two-point Green's functions, it is  by now well-established that, in the Landau gauge, the  gluon propagator  and the ghost dressing function 
are finite in the IR~\cite{Aguilar:2008xm,Boucaud:2008ji}. The finiteness
of both quantities are associated to the phenomenon of dynamical gluon mass generation~\cite{Cornwall:1981zr,Aguilar:2004sw, Aguilar:2006gr, Binosi:2007pi, Aguilar:2011yb,Aguilar:2008xm}.

Although, we have a qualitative understanding of the origin of the finiteness of 
the ghost dressing function, $F(p^2)$, it has been more difficult to obtain from a self-consistent SDE analysis its  entire shape and size  provided by the lattice~\cite{Aguilar:2008xm,Dudal:2012zx,Boucaud:2011eh}.

\begin{figure}[!ht]
\begin{center}
\includegraphics[scale=0.5]{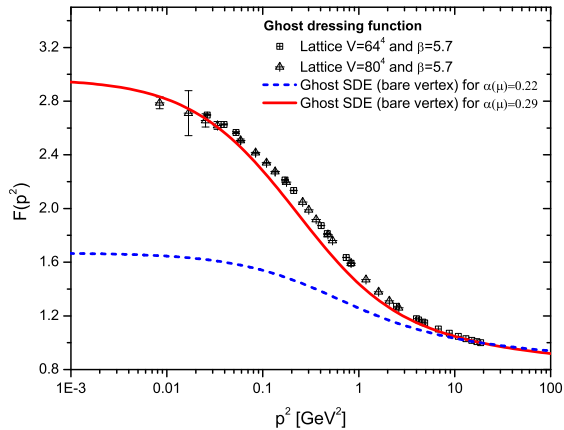}
\caption{Comparison of the ghost dressing function, $F(p^2)$, obtained as  
solution of the ghost SDE when the ghost-gluon vertex is approximate by its bare value, with the
lattice data of Ref.~\cite{Bogolubsky:2007ud}.
The (red) continuous curve represents the case when  $\alpha_s(4.3\, \rm GeV)=0.29$ whereas
the (blue) dashed curve is obtained when  $\alpha_s( 4.3 \,\rm GeV )=0.22$.}
\label{oldsol}
\end{center}
\end{figure}

More specifically, when we substitute the gluon propagator 
furnished by the lattice into the 
ghost SDE, but keeping the  ghost-gluon  vertex at its tree-level value, the resulting $F(p^2)$ is significantly suppressed compared to that of the lattice~\cite{Aguilar:2008xm}; to reproduce the lattice result, one has to artificially  increase the value of the gauge coupling 
from the correct value $\alpha_s = 0.22$ to $\alpha_s = 0.29$~\cite{Aguilar:2010gm}
as illustrated in the Fig.~\ref{oldsol}. 

Given the simple structure of the ghost SDE, it is become clear that the  
origin of the  observed discrepancy is due to the tree-level approximation
we impose for the  fully dressed ghost-gluon vertex, $\Gamma_{\nu}$~\cite{Aguilar:2013xqa}.

Even though preliminary lattice studies indicate that 
the deviations of $\Gamma_{\nu}$ from 
its tree-level value are relatively moderate~\cite{Cucchieri:2004sq,Ilgenfritz:2006he,Sternbeck:2006rd,Cucchieri:2008qm},  it is natural 
to expect that, due to the nonlinear nature of SDE, that even a small deviation can generate significant effects on the kernel of the equation that describes the behavior of the ghost dressing function. Similar studies on the influence of the  three-point functions on the propagators can be found in Refs.~\cite{Dudal:2012zx,Huber:2012kd,Blum:2014gna, Huber:2014bba}.

The main purpose of this talk is present the ``one-loop dressed'' version of the SDE satisfied by one of the form factors appearing in the definition of ghost-gluon vertex, $\Gamma_{\nu}$, in the Landau gauge. 

To be more specific, the tensorial decomposition of  $\Gamma_{\nu}$ consists of two form factors;
however, due to the transversality of the gluon propagator  present
in the ghost SDE, written in Landau gauge,  only the form factor $A(-k,-p,r)$ 
of the ghost momentum $p_{\nu}$ survives.

Here we  present our results for  $A(-k,-p,r)$ in two particular kinematic 
configurations: (i) soft gluon ($k \to 0$) and  (ii) soft ghost ($p \to 0$). 
For the first case, we  compare our results with the lattice data of Ref.~\cite{Ilgenfritz:2006he,Sternbeck:2006rd}. The second configuration is the
relevant one for the ghost SDE. Therefore, we solve numerically the coupled system of integral equations formed by $F(p^2)$ and $A(-k,0,k)$. The numerical solution obtained gives rise to a ghost dressing function that is in excellent agreement with the 
lattice data~\cite{Bogolubsky:2007ud} using the standard value of 
the gauge coupling constant  \mbox{$\alpha_s = 0.22$} which corresponds to 
the momentum-subtraction (MOM) value for 
the point \mbox{$\mu=4.3$ GeV}~\cite{Boucaud:2008gn}, used 
to renormalize the gluon propagator obtained from the lattice~\cite{Aguilar:2013xqa}.

\section{The ghost SDE and the ghost-gluon vertex}

\begin{figure}[h]
\begin{center}
\includegraphics[scale=0.85]{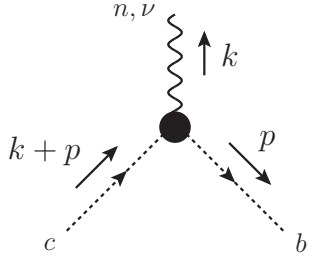}
\caption{The fully dressed ghost-gluon vertex.}
\label{fullgg}
\end{center}
\end{figure}

 We start by denoting  the full ghost-gluon vertex, shown in Fig.~\ref{fullgg}, by 
\bea
\Gamma_{\nu}^{nbc}(-k,-p,r) = gf^{nbc}\Gamma_{\nu}(-k,-p,r)\,, \quad r=k+p\,,
\label{Gdef}
\eea
with $k$ representing the momentum of the gluon and $p$ of the anti-ghost. 

The most general Lorentz structure of this vertex is given by
\be
{\Gamma}_{\nu}(-k,-p,r) =  A(-k,-p,r)\, p_{\nu} + B(-k,-p,r) \,k_{\nu} \,.
\label{Gtens}
\ee 
 
At tree-level, the two form factors assume the values 
\mbox{$A^{[0]}(-k,-p,r)=1$} and \mbox{$B^{[0]}(-k,-p,r)=0$},  giving rise to the well know bare ghost-gluon vertex  
$\Gamma^{[0]}_{\nu}=p_\nu$.

It is convenient to define now the  so-called ``Taylor limit'' of the ghost-gluon vertex, where the vertex has vanishing ghost momentum, $r=0$, $p=-k$.
In this special kinematic configuration, the $\Gamma_\nu (-k,-p,r)$ of Eq.~(\ref{Gtens}) becomes
\be
\Gamma_\nu(-k,k,0) = -[A(-k,k,0) - B(-k,k,0)]k_\nu \,.
\label{GtensTayl}
\ee

Intimately connected to this limit is the well-known Taylor theorem, which states that, to all orders in perturbation theory~ \cite{Taylor:1971ff}. ,
\be
A(-k,k,0) - B(-k,k,0) = 1;
\label{Taylteor}
\ee
as a result, the fully-dressed vertex assumes the tree-level value corresponding to this particular kinematic configuration, \ie $\Gamma_\nu(-k,k,0)=-k_\nu$.

\begin{figure}[!t]
\includegraphics[scale=0.7]{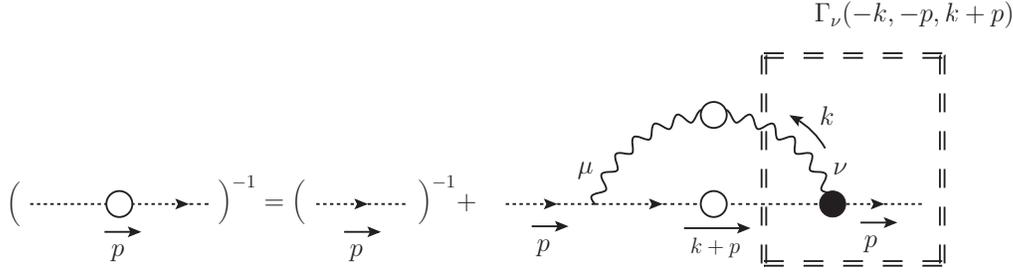}
\caption{The SDE for the ghost propagator given by Eq.~(2.5). The white blobs represent the
fully dressed gluon and ghost propagators, while the black blob denotes the dressed ghost-gluon
vertex.}
\label{ghostSDE}
\end{figure}

Of extreme importance for this analysis is to understand how different
behaviors for the ghost-gluon vertex affects the structure of the ghost SDE,
diagrammatically represented in the Fig.~\ref{ghostSDE}, and  written as
\be
iD^{-1}(p^2) = ip^2 - g^2 C_{\rm {A}}  \int_k
\Gamma_{\mu}^{[0]}(k,-k-p,p)\Delta^{\mu\nu}(k)\Gamma_{\nu}(-k,-p,k+p) D(k+p)\,,
\label{SDgh}
\ee
where $C_{\rm A}$ denotes the Casimir eigenvalue of the adjoint representation ($N$ for $SU(N)$), 
$d=4-\epsilon$ is the space-time dimension, and we have introduced the integral measure
\be
\int_{k}\equiv\frac{\mu^{\epsilon}}{(2\pi)^{d}}\!\int\!\mathrm{d}^d k,
\label{dqd}
\ee
with $\mu$ the 't Hooft mass.
In the Landau gauge, the gluon
propagator $\Delta_{\mu\nu}(q)$ has the transverse form
\be 
\Delta_{\mu\nu}(q)=-i P_{\mu\nu}(q)\Delta(q^2)\, \quad \mbox{with} \quad P_{\mu\nu}(q)= g_{\mu\nu} - \frac{q_\mu q_\nu}{q^2} \,. 
\label{prop_cov}
\ee

Due to the full transversality of $\Delta_{\mu\nu}(k)$, 
any reference to the form factor $B$  
disappears from the ghost SDE of \1eq{SDgh}. Specifically, 
substituting \1eq{Gtens} into \1eq{SDgh}, and 
introducing the usual renormalization constants we obtain~\cite{Aguilar:2013xqa} 
\be
F^{-1}(p^2) = Z_c +ig^2 C_{\rm {A}} \int_k\, \left[1-\frac{(k\cdot p)^2}{k^2p^2}\right] A(-k,-p,k+p)\Delta (k)  D(k+p) \,,
\label{tt2r}
\ee
where we have  introduced the ghost dressing function, $F(q^2)$, defined as 
$F(q^2)= q^2D(q^2)$.

Notice that the closed expression of $Z_c$ is obtained from \1eq{tt2r} itself, by imposing the MOM renormalization condition on $F(q^2)$,  {\it i.e} 
\mbox{$F(q^2=\mu^2) =1$}, where $\mu$ is the renormalization point where the 
dressing function assumes its tree-level value.

Evidently, \1eq{tt2r} couples the unknown functions $A(-k,-p,k+p)$, described by the corresponding vertex SDE, and $F(p^2)$.  Given that $A(-k,-p,k+p)$ is a function of three variables, 
$p^2$, $k^2$, and the angle between the two (appearing in the inner product $p\cdot k$), 
a full SDE treatment is rather cumbersome, and lies beyond our present technical powers. Instead, we will consider the behavior of $A(-k,-p,k+p)$ for vanishing $p$; to that end, 
we start out with the Taylor expansion of $A(-k,-p,k+p)$ around $p=0$,  
and we only keep the first term, $A(-k,0,k)$, thus converting $A$ into a function of a single variable. We emphasize that the limit $p\to 0$ is taken only inside the argument of the form factor $A$, but not in the rest 
of the terms appearing in the SDE of \1eq{tt2r}~\cite{Aguilar:2013xqa}.
 
Thus, the approximate version of the SDE in \1eq{tt2r} reads
\be
F^{-1}(p^2) = Z_c +ig^2 C_{\rm {A}} \int_k\, \left[1-\frac{(k\cdot p)^2}{k^2p^2}\right] A(-k,0,k)\Delta (k)  D(k+p) \,.
\label{tt2app}
\ee

\section{The one-loop dressed approximation for the vertex}\label{sec.vertexSDE}

In this section, we will schematize  the derivation of the nonperturbative expression for the form factor $A$,  
in two special kinematic configurations: (i) the \emph{soft gluon limit}, in which the momentum 
carried by the gluon leg is zero ($k=0$), and (ii) the \emph{soft ghost limit}, 
where the momentum of the anti-ghost leg vanishes ($p=0$). 

\begin{figure}[!ht]
\includegraphics[width=16cm]{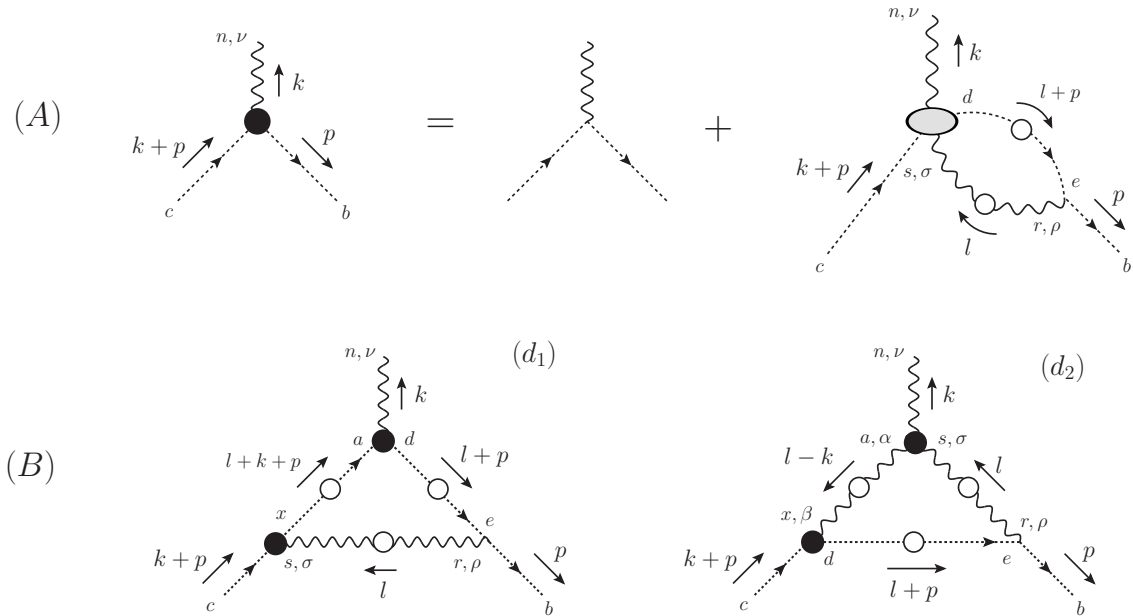}
\caption{(A) The complete SDE of the ghost-gluon vertex. (B) The diagrams that contributes to the ghost-gluon vertex SDE after the ``one-loop dressed'' expansion of the ghost-gluon kernel.}
\label{vertexSDE}
\end{figure}

To do that, we start by showing in  Fig.~\ref{vertexSDE} the diagrammatic representation of the SDE satisfied by the ghost-gluon vertex. Besides the full gluon and ghost propagators, we observe  
that the most relevant quantity, which controls the dynamics of this vertex, is the four-point ghost-gluon kernel represented by the gray blob.  

In order to derive a SDE for the ghost-gluon vertex tractable, we will replace
the  ghost-gluon kernel by its ``one-loop dressed'' approximation~\cite{Schleifenbaum:2004id}. In other words, the skeleton expansion of the four-point ghost-gluon kernel will only include 
the diagrams appearing in panel $(B)$ of Fig.~\ref{vertexSDE}.
 
Thus, the approximate version 
of the SDE that we employ may be cast in the form
\begin{equation}\label{glghSDE}
\Gamma_\nu(-k,-p,k+p) = p_\nu - \frac{i}{2}g^2 C_{\rm A} [(d_1)_\nu - (d_2)_\nu]\,,
\end{equation}
where the diagrams $(d_i)$ are given by
\begin{eqnarray}\label{SDEdiagrams}
(d_1)_\nu &=& \int_l \Gamma_\rho^{[0]}\Delta^{\rho\sigma}(l)\Gamma_\sigma D(l+k+p) \Gamma_\nu D(l+p)\,, \nonumber \\
(d_2)_\nu &=& \int_l \Gamma_\rho^{[0]}\Delta^{\rho\sigma}(l)\Gamma_{\nu\sigma\alpha}\Delta^{\alpha\beta}(l-k)\Gamma_\beta D(l+p)\,.
\end{eqnarray}

Moreover, in the two diagrams, $(d_1)$ and $(d_2)$,  we will keep fully dressed propagators, but we will 
replace the fully dressed three-gluon vertex appearing in graph $(d_2)$ 
by its tree-level expression, namely 
\begin{equation}
\label{treeglvertex}
\Gamma_{\alpha\mu\nu}(q,r,p) \rightarrow
\Gamma_{\alpha\mu\nu}^{[0]}(q,r,p) = (r-p)_\alpha g_{\mu\nu} + (p-q)_\mu g_{\nu\alpha} + (q-r)_\nu g_{\alpha\mu}\,.
\end{equation}

It is important to notice the presence of the fully dressed ghost-gluon vertices, in the diagrams, $(d_1)$ and $(d_2)$. In the sequence, we will mention  
the   additional approximations will be imposed on those vertices, 
depending on the specific details of each kinematic case considered.

\subsection{\label{softh}Soft gluon configuration}

The first kinematic configuration we will analyse is the so-called
\emph{soft gluon limit}, where the gluon leg has a  vanishing momentum, $k=0$.
In this case the ghost-gluon vertex becomes a function of only one momentum, $p$, and
it is entirely expressed  in terms of a single form factor, namely,
\begin{equation}\label{vertexA}
\Gamma_\nu(0,-p,p) = A(p) p_\nu; \quad A(p)\equiv A(0,-p,p)\,.
\end{equation}

Setting $k=0$ in Eq.~(\ref{glghSDE}), one is able to isolate the form factor $A$ by means of the projection
\begin{equation}\label{factorA}
A(p) = 1 - \frac{i}{2}g^2 C_A [(d_1) - (d_2)]; \quad (d_i) \equiv \frac{p^\nu}{p^2}(d_i)_\nu\,, \quad i=1,2\,,
\end{equation}
where the diagrams $(d_i)$ are obtained from those of Eq.~(\ref{SDEdiagrams}) in the limit $k\rightarrow 0$.

Notice that in the limit $k=0$, 
the  vertex $\Gamma_{\nu}$ entering in graph $(d_1)$ becomes $\Gamma_{\nu}(0, -l-p,l+p)$, which allows one to derive 
a linear integral equation for the unknown quantity  for $A(0,-p,p)$.

On the other hand, the same  momenta configuration
does not happen to the remaining 
ghost-gluon vertices, namely, $\Gamma_\sigma$ and $\Gamma_\beta$ in graphs $(d_1)$ and $(d_2)$, 
respectively; their arguments depend on all possible 
kinematic variables, and the inclusion of the full $A$ 
would give rise to a (non-linear) integral equation, too complicated to solve. 
We therefore approximate all remaining ghost-gluon vertices by their tree-level 
expressions.

Thus, using the notation introduced in \1eq{factorA}, the 
diagram $(d_1)$ reads 
\begin{equation}\label{d1projection}
(d_1) = \int_l \frac{(l\cdot p)}{(l+p)^2 p^2}[(l\cdot p)^2 - l^2 p^2] D^2(l)\Delta(l+p)A(l)\,;
\end{equation}
while  $(d_2)$ is given by 
\begin{equation}\label{d2projection}
(d_2) = 2 \int_l \frac{(l\cdot p)}{l^2 p^2}[l^2 p^2 - (l\cdot p)^2] \Delta^2(l) D(l+p)\,.
\end{equation}

Combining the above expressions, the final answer for $A(p)$, in Euclidean space, is given by~\cite{Aguilar:2013xqa}
\begin{eqnarray}\label{Aeuclidean}
A(x) &=& 1 - \frac{\alpha_s C_A}{4\pi^2}\int_0^\infty \!\!\! dt \sqrt{xt}\,F^2(t) A(t) \int_0^\pi \!\!\!d\theta\sin^4\theta\cos\theta\left[\frac{\Delta(z)}{z}\right] \nonumber \\
&-& \frac{\alpha_s C_A}{2\pi^2}\int_0^\infty \!\!\! dt \sqrt{xt}\, t\,\Delta^2(t) \int_0^\pi \!\!\! d\theta\sin^4\theta\cos\theta\left[\frac{F(z)}{z}\right] \,,
\end{eqnarray}
with  \mbox{$l^2=t$};  \mbox{$p^2=x$}; \mbox{$(l+p)^2=z$}; \mbox{$(l\cdot p) = \sqrt{xt}\cos\theta$}. In addition, we have used $g^2=4\pi \alpha_s$, and re-express the ghost propagators in terms of their dressing functions.

It is interesting to notice that, when the momentum of the ghost leg is also zero, {\it i.e.} $x=0$, Eq.~(\ref{Aeuclidean}) reproduces the tree-level value of the form factor, \textit{i.e.}, $A(0)=1$.

\subsection{Soft ghost configuration (Taylor kinematics)}

Now, let us  derive an approximate version for $A$ in the soft ghost configuration, to be denoted by
\begin{equation}\label{Asoftghost}
\lim_{p\rightarrow 0}A(-k,-p,k+p) = A(-k,0,k) \equiv A(k)\,.
\end{equation}

As mentioned before, we will employ the soft ghost  configuration into the ghost
SDE and check the improvements, this configuration may make in the description of the IR behavior of the ghost dressing function.

It is important to emphasize here that the form factor $A(-k,0,k)$ 
obtained in the soft ghost configuration  coincides with that of the Taylor kinematics $A(-k,k,0)$, appearing  in the constraint imposed by Taylor's theorem, given by Eq.~(\ref{Taylteor}). A detailed proof of the equivalence of $A(-k,0,k)=A(-k,k,0)$ can be found in Ref.~\cite{Aguilar:2013xqa}.

Let us now derive  the explicit expression for the form factor $A$ in the soft ghost limit. Again, your starting point are the  diagrams shown in panel $(B)$ of Fig.~\ref{vertexSDE},  where we  dress up all  gluon and ghost propagators, and keep tree-level values for \textit{all} the interaction vertices. 

In this configuration, the expressions given in Eq.~(\ref{SDEdiagrams}) reduce to
\begin{eqnarray}\label{diagramsp0}
(d_1)_\nu &=& p^\rho (k+p)^\sigma \int_l (l+p)_\nu D(l+p)D(l+k+p)\Delta(l)P_{\rho\sigma}(l)\,, \nonumber \\
(d_2)_\nu &=& p^\rho (k+p)^\beta \int_l D(l+p)\Delta(l)\Delta(l-k)P_\rho^\sigma(l)P^\alpha_\beta(l-k)\Gamma^{[0]}_{\nu\sigma\alpha}\,,
\end{eqnarray}
 
 The general procedure for isolating the $A(-k,0,k)$ from the diagrams $(d_1)$ and $(d_2)$, must be done with care.  Here, due to space limitations,  we will only outline the basic steps that should be performed: ({\it i}) Set $p=0$ 
from the beginning \textit{inside} the integrals of Eq.~(\ref{diagramsp0}); ({\it ii}) Discard all the terms that give rise to structures 
of the type ${\cal O}(p)(k+p)_\nu$; 
({\it iii}) Determine the contribution of the diagram that saturates the index of the momentum $p^\rho$ with the metric tensor $g_{\nu\rho}$. For more details see Ref.~\cite{Aguilar:2013xqa}.

After performing the steps described above, 
we arrive at the final result (in Euclidean space)
\begin{eqnarray}\label{Ap01}
A(y) &=& 1 - \frac{\alpha_s C_{\rm A}}{12\pi^2}\int_0^\infty \!\!\!dt \,\sqrt{yt}\,F(t)\Delta(t) \int_0^\pi \!\!\!d\theta'\sin^4\theta'\cos\theta'\left[\frac{F(u)}{u}\right]  \\
&+&\frac{\alpha_s C_{\rm A}}{6\pi^2} \int_0^\infty \!\!\! dt \,F(t)\Delta(t) \int_0^\pi \!\!\! d\theta' \sin^4\theta' \left[ \frac{\Delta(u)}{u}\right][yt(1+\sin^2\theta')-(y+t)\sqrt{yt}\cos\theta'] \nonumber \,.
\end{eqnarray}
where, in this case, $y=k^2$, $u = (l + k)^2$, and $\theta'$ is the angle between $k$ and $l$.

\section{\label{numsoftgluon}Numerical result for the soft gluon configuration}

\begin{figure}[!t]
\begin{minipage}[b]{0.45\linewidth}
\centering
\includegraphics[scale=0.50]{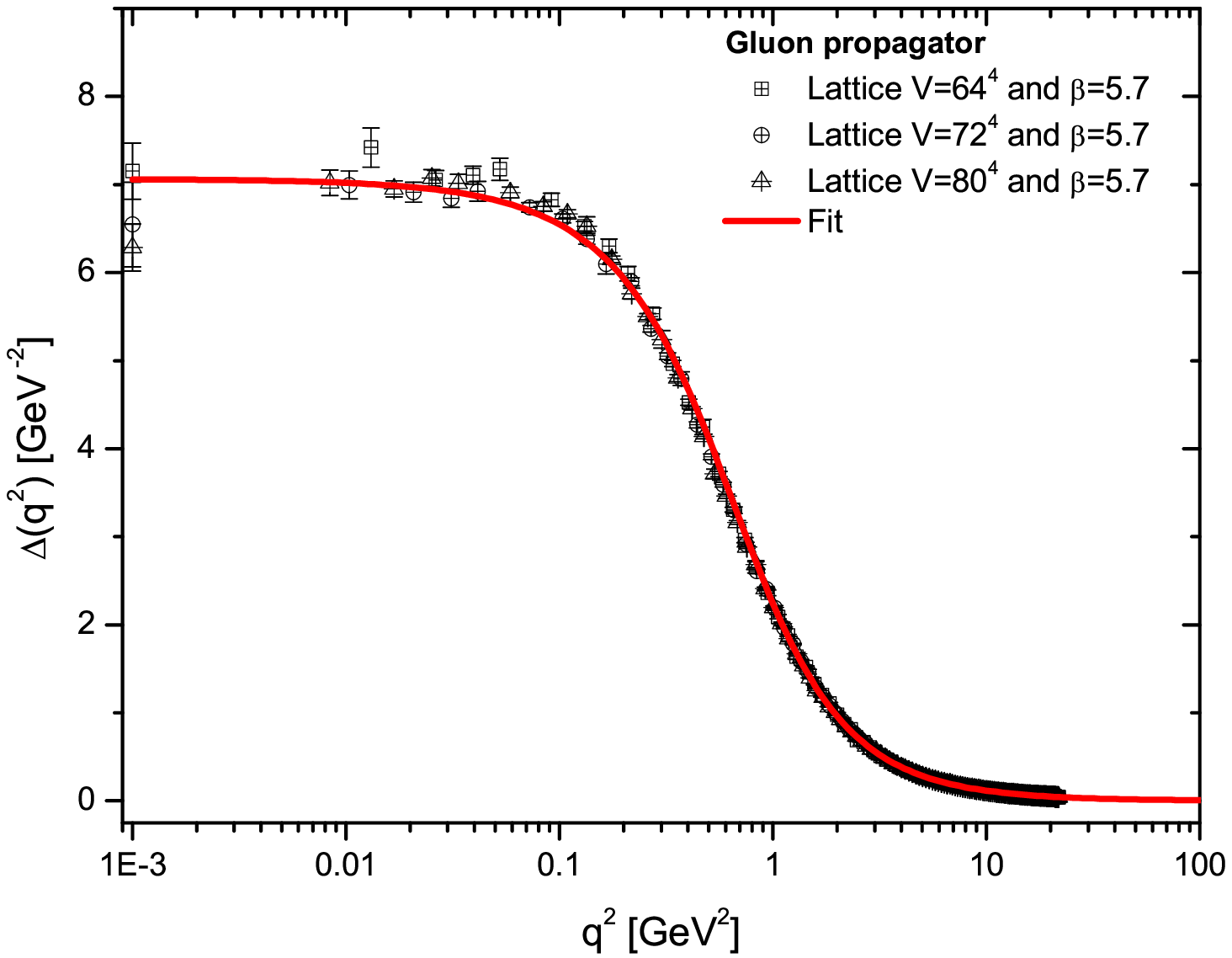}
\end{minipage}
\hspace{0.5cm}
\begin{minipage}[b]{0.50\linewidth}
\includegraphics[scale=0.50]{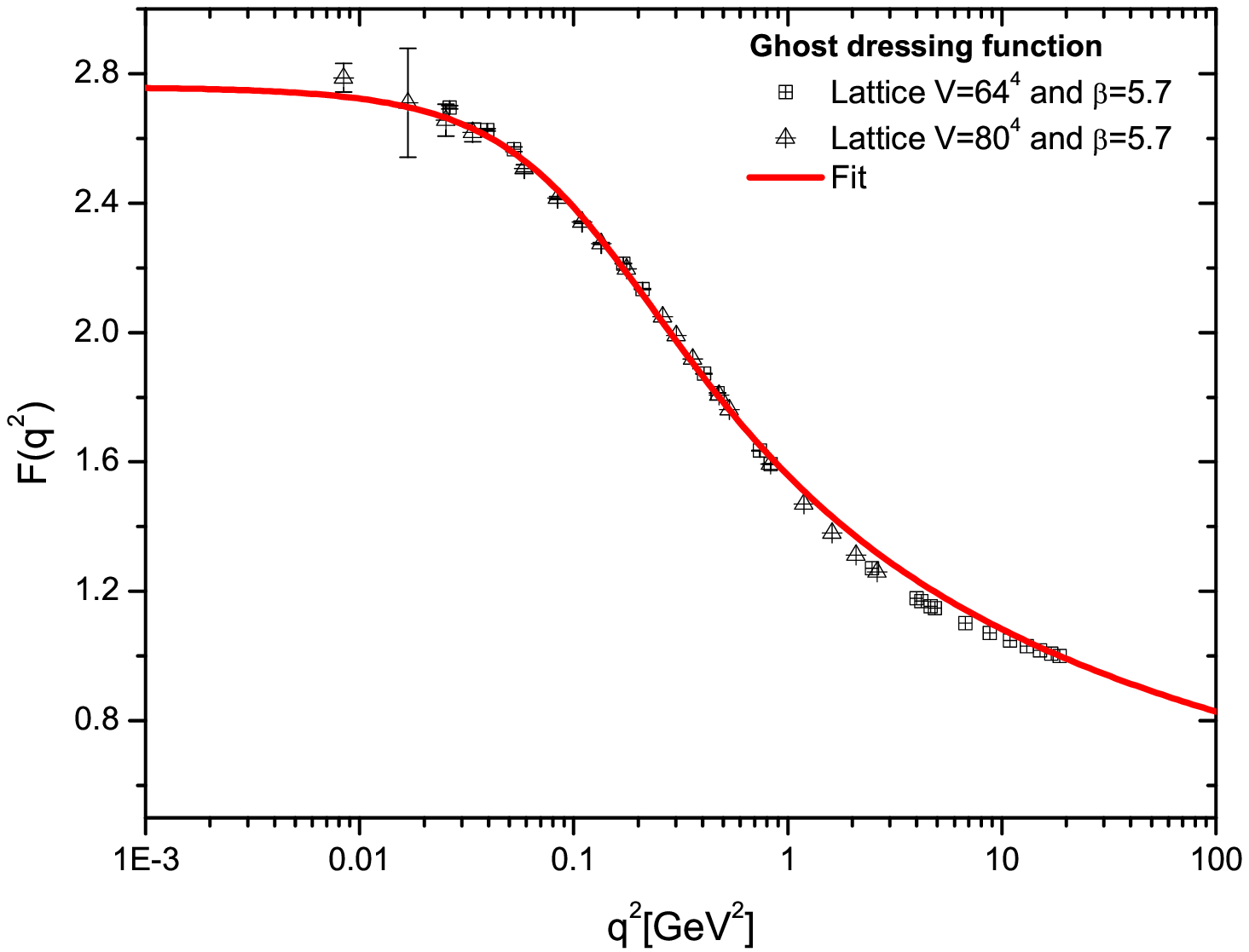}
\end{minipage}
\caption{Lattice results for the gluon propagator, $\Delta(q)$, (left panel) and ghost dressing, $F(q)$, (right panel)
obtained in  Ref.~\cite{Bogolubsky:2007ud} and renormalized at $\mu=4.3$ GeV.}
\label{fig1}
\end{figure}

Now we are in position to solve numerically the equations obtained in the previous sections.  More specifically, in this section we determine $A(0,-p,p)$ by solving the integral equation (\ref{Aeuclidean}) through an iterative process. To do that, we  use as input the  lattice data 
obtained from the $SU(3)$ quenched simulations of~\cite{Bogolubsky:2007ud}, renormalized at $\mu=4.3$ GeV, within the MOM scheme (see Fig.~\ref{fig1}).

In addition, we will employ $\alpha_s(\mu)=0.22$ that is the 
standard value for the gauge coupling obtained from the higher-order calculation presented in~\cite{Boucaud:2008gn}. 

\begin{figure}[!ht]
\begin{center}
\includegraphics[scale=0.5]{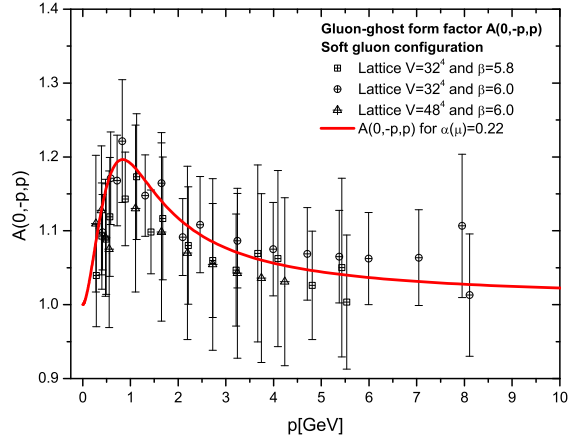}
\caption{ Comparison of our numerical result for $A(0,-p,p)$, obtained from Eq.(3.8) when \mbox{$\alpha_s(\mu)=0.22$} (red line) and the lattice data of Ref.~\cite{Ilgenfritz:2006he,Sternbeck:2006rd}.}
\label{fig2}
\end{center}
\end{figure}

Our result for the soft gluon configuration are presented in the Fig.~\ref{fig2}, where the (red) curve represents the corresponding solution for $A(0,-p,p)$. Notice that $A(0,-p,p)$ 
develops a considerable peak around the momentum region of
\mbox{$830$ MeV}.  It is also
interesting to notice that  in both,  IR and ultraviolet limits, the form factor $A$ gradually
approaches its tree level value. 

In the same Fig.~\ref{fig2}, we compare our numerical results with the  
corresponding lattice data obtained
in Ref.~\cite{Ilgenfritz:2006he,Sternbeck:2006rd} for this particular kinematic configuration. 

Although, the error bars are rather pronounced, we clearly see that 
our solution follows the general structure of
the data. In particular, notice that both 
peaks occur in the same intermediate region of momenta, where $A(0,-p,p)$ receives a significant non-perturbative correction, 
deviating considerably from its tree level value. 

\section{ \label{system}  Numerical results for the coupled system}

In this section, we solve numerically the coupled system formed by the integral equations of the ghost dressing function (\ref{tt2app}) and the one for the  ghost-gluon vertex in the soft ghost configuration, given by (\ref{Ap01}). 
Here again, the unique external ingredient used when solving this system are the  lattice data for the gluon propagator $\Delta(q)$ shown on the left panel of Fig.~ \ref{fig1}. In particular, we are interested in analysing how the inclusion of a non-trivial corrections for the corresponding ghost-gluon vertex induces modifications in the ghost dressing function.

\begin{figure}[!t]
\begin{minipage}[b]{0.45\linewidth}
\centering
\includegraphics[scale=0.50]{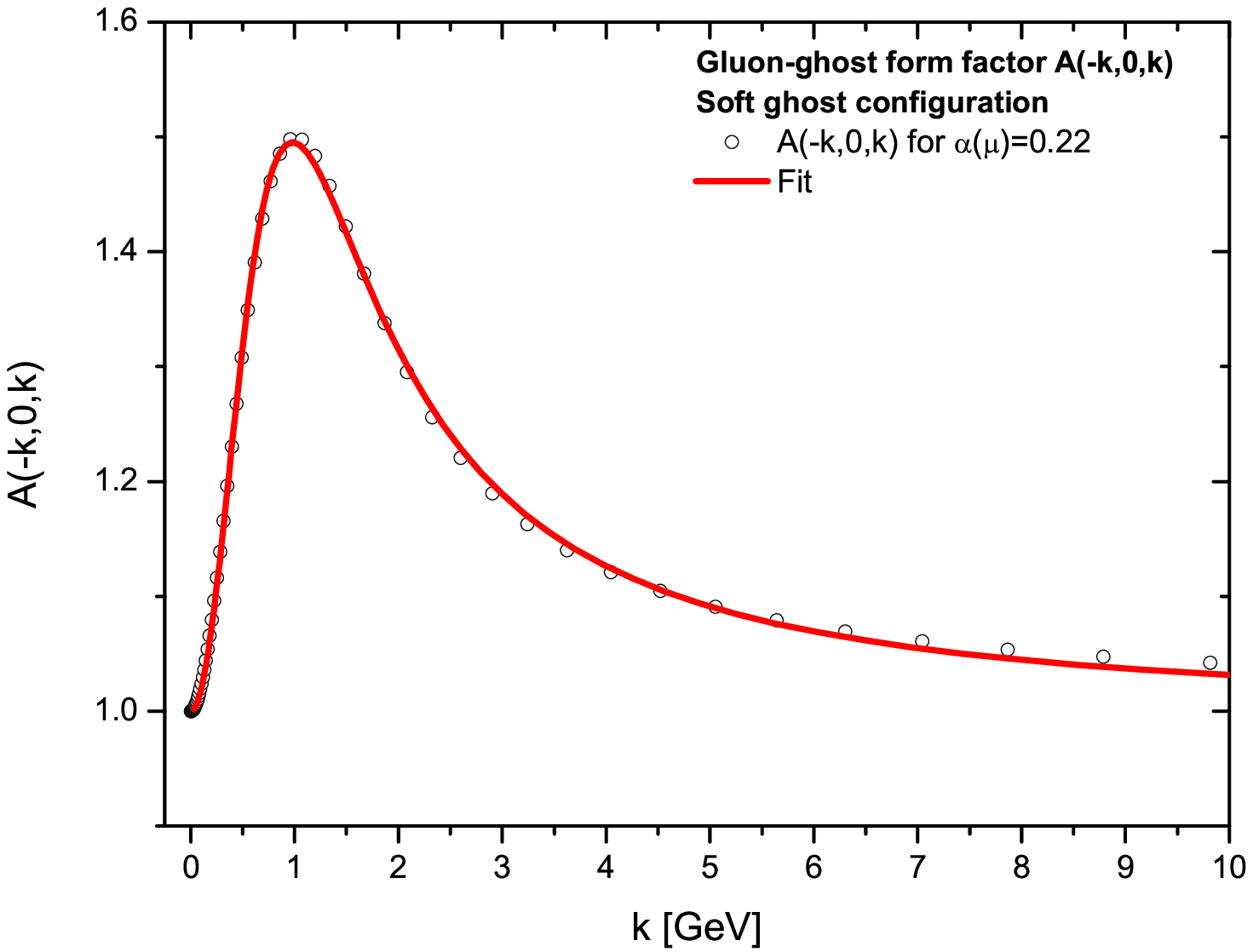}
\end{minipage}
\hspace{0.5cm}
\begin{minipage}[b]{0.50\linewidth}
\includegraphics[scale=0.50]{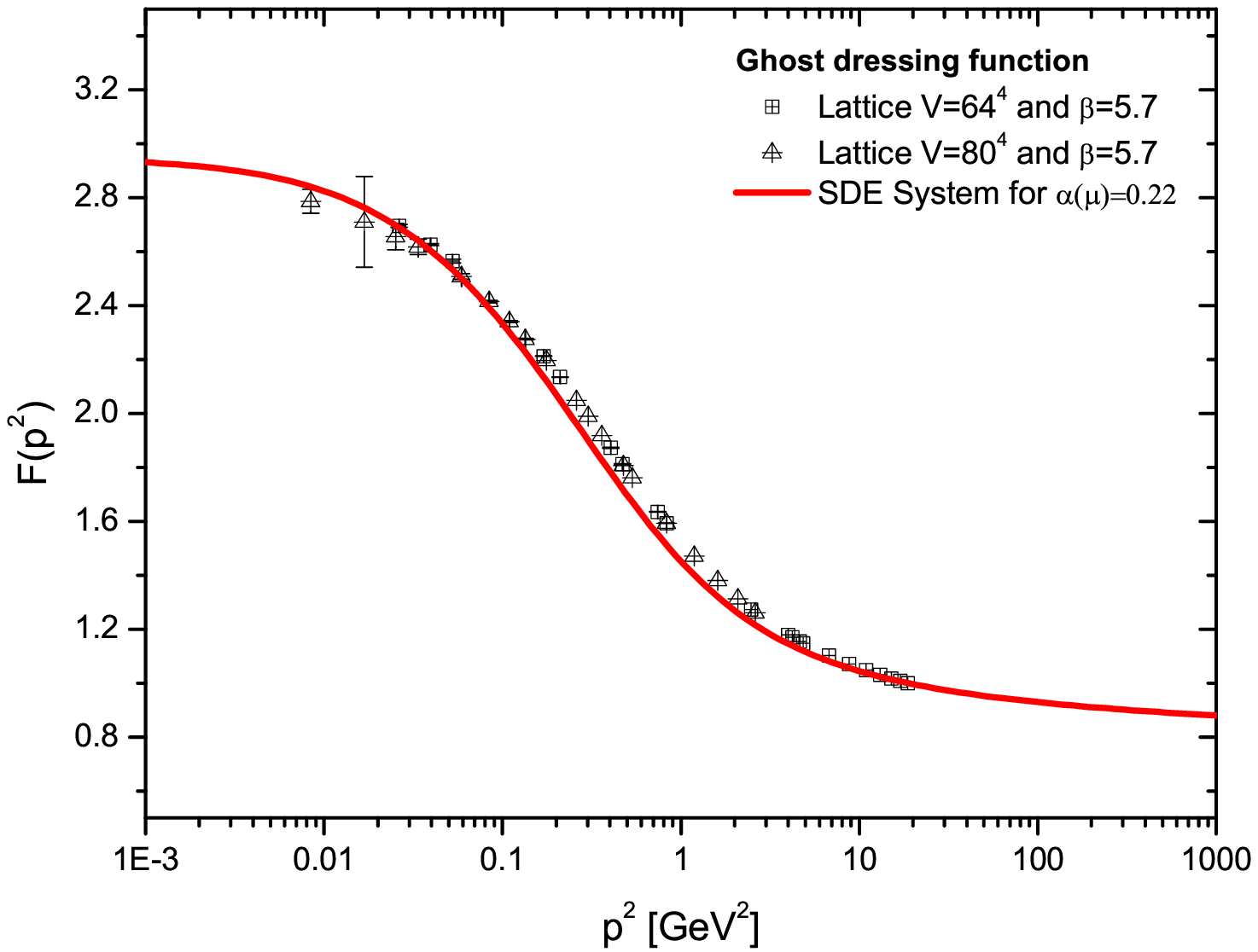}
\end{minipage}
\caption{ {\it Left panel}: The form factor $A(-k,0,k)$ (circles) and
the fit given by Eq.(4.1) (red continuous line).  
{\it Right panel}: The numerical solution of $F(p)$ (red continuous line) compared  with the lattice data of Ref.~\cite{Bogolubsky:2007ud}. Note that the value of $\alpha_s$ used when solving the system is $\alpha_s(\mu)  =0.22$.}
\label{fig3}
\end{figure}

The results for $F(p)$ and  $A(-k,0,k)$ when $\alpha_s(\mu)=0.22$ are shown in~Fig.~\ref{fig3}. On the left panel, the circles represents our result for $A(-k,0,k)$.    Here again, $A(-k,0,k)$ displays the same pattern  found in the soft gluon configuration. More specifically, the peak reaches its maximum around \mbox{$1$ GeV} and the curve tends to its tree-level value in the IR and ultraviolet limits. Unfortunately, in this plot we do not compare our results with the lattice predictions, since  there are no available lattice data for the ghost soft configuration.

 The (red) continuous curve of~Fig.~\ref{fig3}, represents the fit for $A(-k,0,k)$, whose functional form is given by
\be
A(-k,0,k)= 1 + \frac{ak^2}{[(k^2+b)^2 + c]}\ln\big(d + k^2/k_0^2\big) \,,
\label{fit}
\ee
where the adjustable parameters are  \mbox{$a=0.68\,\mbox{GeV}^2$}, \mbox{$b=0.72\,\mbox{GeV}^2$}, \mbox{$c=0.29\,\mbox{GeV}^4$}, \mbox{$d=9.62$} and \mbox{$k_0^2=1\,\mbox{GeV}^2$}.

The comparison of $F(p)$ (red continuous curve) obtained as solution 
from the system of Eqs.~(\ref{tt2app}) and (\ref{Ap01}) with the  
lattice data of Ref.~\cite{Bogolubsky:2007ud} are shown
on the right panel of~Fig.~\ref{fig3}. As we can clearly see,
we find a remarkable agreement between the curves, even keeping the standard value of the coupling constant, {\it i.e.} $\alpha_s(\mu)=0.22$.

From the results presented in~Fig.~\ref{fig3}, we can conclude that although $A(-k,0,k)$ is not very different from its value at tree level in the deep IR region, however,  the presence of the  peak, in its intermediate region, when integrated in the kernel of ghost SDE is sufficient for  increasing the saturation point from $F(0)=1.67$ to $F(0)= 2.95$ (see Figs.~\ref{oldsol} and \ref{fig3}, respectively).

This observation suggests that the ghost SDE is particularly sensitive to the values of its ingredients  at momenta around two to three times the  QCD mass scale.

For the sake of completeness, we will repeat the same analysis  for the $SU(2)$ gauge group. To do that, we set  $C_{\rm A}=2$ in the system of  Eqs.~(\ref{tt2app}) and
 (\ref{Ap01}), and  we solve the system using as  input the $SU(2)$ data for $\Delta(q)$ of Ref.~\cite{Cucchieri:2007md} renormalized at $\mu=2.2$ GeV, moreover we  use the value $\alpha_s(\mu)=0.81$.

\begin{figure}[!t]
\begin{minipage}[b]{0.45\linewidth}
\centering
\includegraphics[scale=0.33]{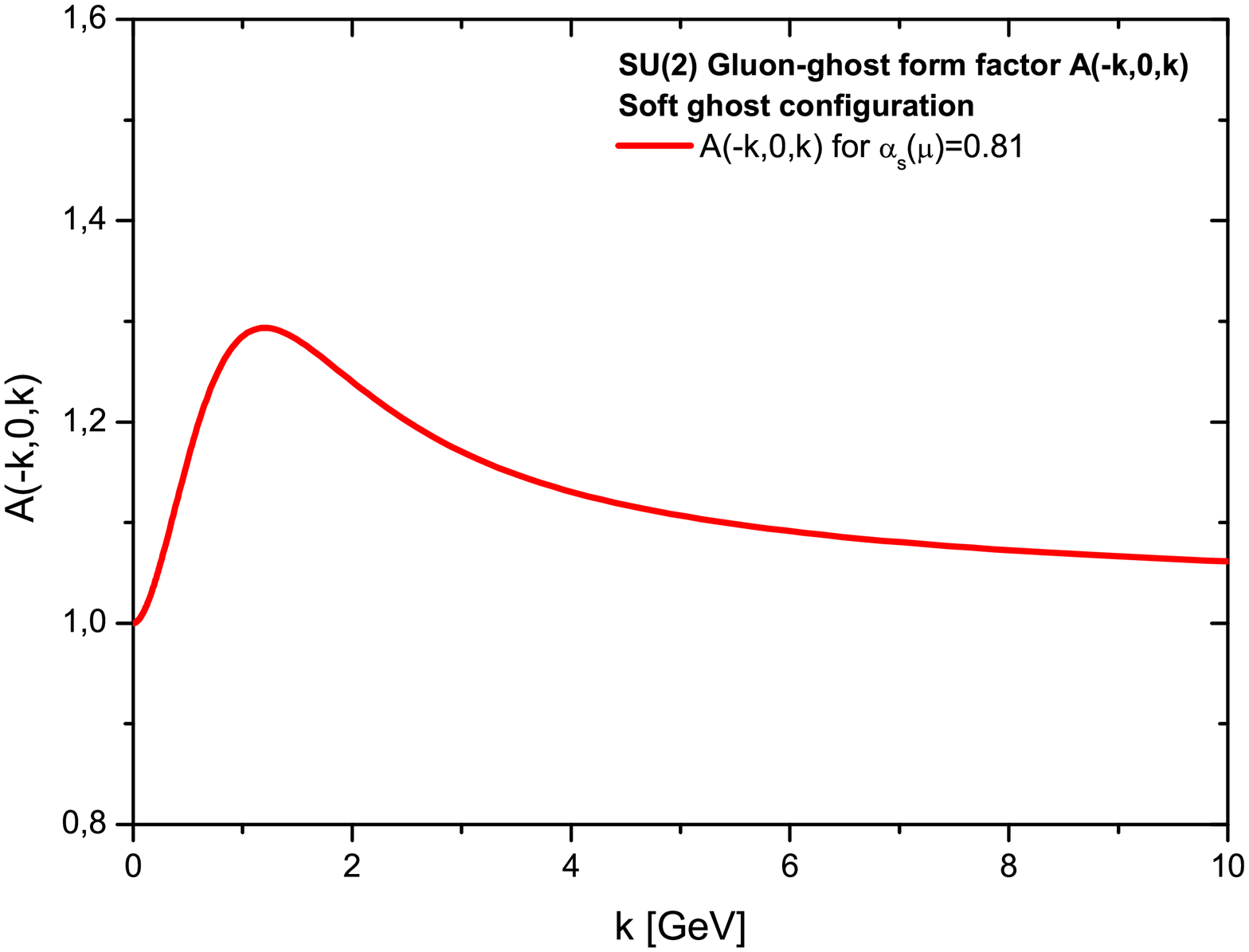}
\end{minipage}
\hspace{0.5cm}
\begin{minipage}[b]{0.50\linewidth}
\includegraphics[scale=0.50]{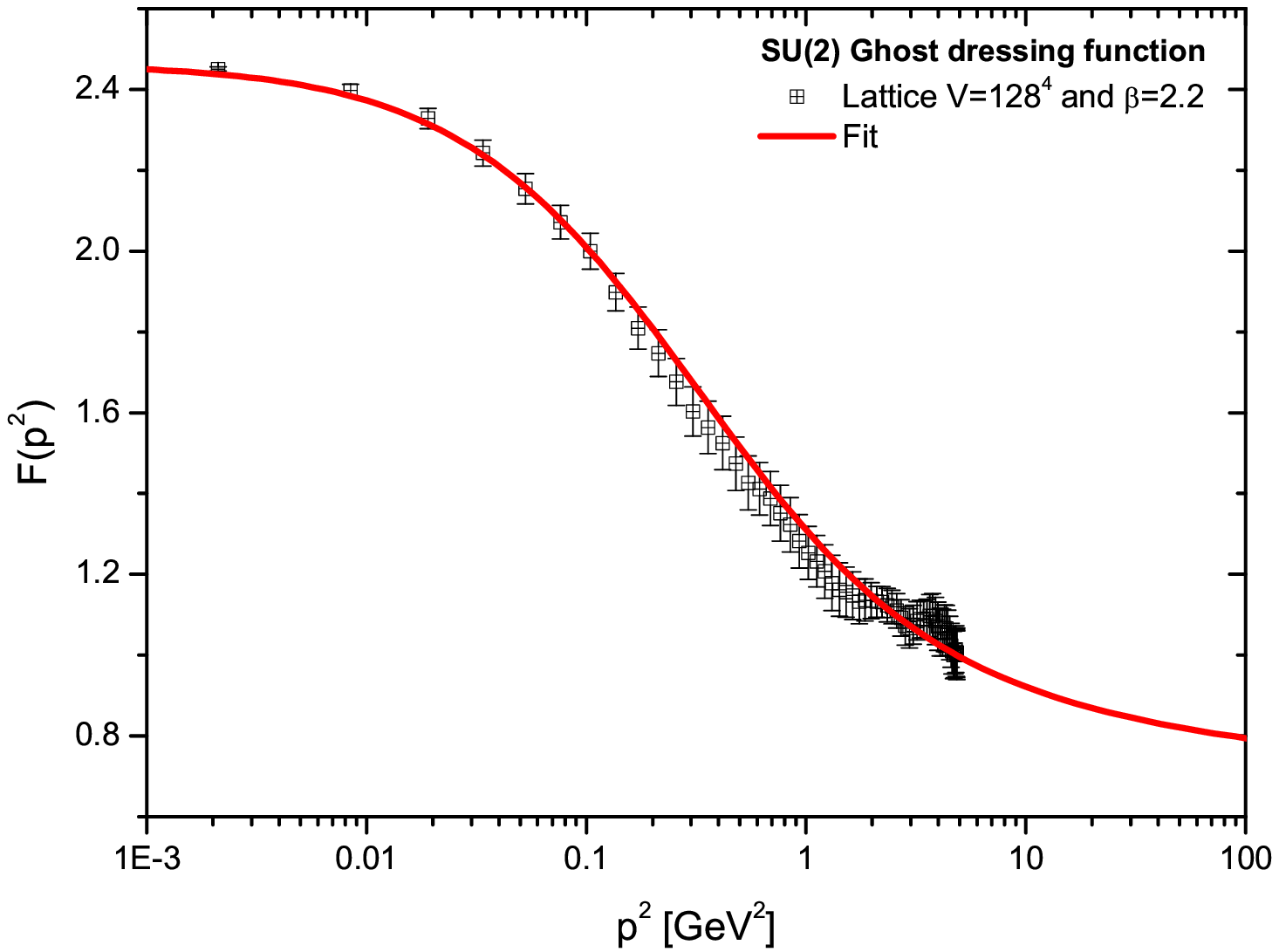}
\end{minipage}
\caption{ 
{\it Left panel}: The numerical result for $A(-k,0,k)$  obtained from the system of Eqs.~(2.9) and (3.11) when $\alpha_s(\mu)=0.81$ (red continuous line). 
{\it Right panel}: Comparison of $F(p)$ (red continuous line) obtained from as solution from the coupled system with
the $SU(2)$ lattice data of Ref.~\cite{Cucchieri:2007md}.}
\label{softgsu2}
\end{figure}
 
On the left panel of~Fig.~\ref{softgsu2}, we show our numerical results for $A(-k,0,k)$. Once again, we see that the  solution displays the same qualitative behavior found in the $SU(3)$ analysis, shown in~Fig.~\ref{fig3}. However, for the $SU(2)$, the peak is slightly shifted towards to the ultraviolet, occurring at around \mbox{$1.2$ GeV}.

The result for  $F(p)$ (red continuous curve) is presented on the right panel of~Fig.~\ref{softgsu2}. We clearly see, once more, the excellent agreement 
with the corresponding lattice data of~\cite{Cucchieri:2007md}. 

Notice that also in the  $SU(2)$ case, 
the introduction of the non-perturbative correction to the ghost-gluon vertex
reduces considerably the value of the gauge coupling needed to reproduce the 
lattice data. Specifically,  when we employ the bare vertex 
the lattice result is reproduced for  $\alpha_s(\mu)=0.99$, whereas the value of the 
coupling used when solving the system of $F(p)$ and $A(-k,0,k)$ is $\alpha_s(\mu)=0.81$.

\section{\label{concl}Conclusions}

In this talk  we have presented 
a study of the impact of the ghost-gluon vertex   
on the overall shape of dressing of the ghost propagator obtained as solution 
of the ghost SDE for different $SU(N)$ gauge groups ($N=2,3$).

To do that, we have focused on the dynamics of the ghost-gluon form factor, denoted by $A$, which survives in the SDE for ghost dressing function, in the Landau gauge.
Using the  ``one-loop dressed'' approximation of the SDE that 
governs the evolution of the ghost-gluon vertex, we have evaluated $A$ in two special kinematic configurations: (i) the soft gluon and  (ii) the soft ghost limits. 

In both limits, the result obtained for $A$ 
displays a reasonable peak around 1 GeV, corresponding to a 20$\%$ and 50$\%$ increase  with respect to the tree-level value, respectively. For the case of the soft gluon configuration, we have also shown that our result   
compares rather well with the existing lattice data~\cite{Ilgenfritz:2006he,Sternbeck:2006rd}.

In addition, we have demonstrated that when the soft ghost  kinematic limit is coupled  to the ghost SDE, the contribution of this particular 
form factor accounts for the missing strength of the associated kernel, 
allowing one to reproduce the lattice results for the ghost dressing function rather accurately, using the standard value of 
the gauge coupling constant. Therefore, we conclude that the ghost SDE is particularly sensitive to the values of its ingredients  at momenta around 1 GeV.

\acknowledgments 
I would like to thank the ECT* for the hospitality and for supporting
the QCD-TNT III organization.
The research  of  A.~C.~A  is supported by the 
National Council for Scientific and Technological Development - CNPq
under the grant 306537/2012-5 and project 473260/2012-3,
and by S\~ao Paulo Research Foundation - FAPESP through the project 2012/15643-1.

\end{document}